\documentclass[12pt]{article}
\usepackage{amsmath}
\usepackage{amssymb}
\DeclareMathAlphabet{\mathpzc}{OT1}{pzc}{m}{it}
\usepackage{algorithm}
\usepackage{algpseudocode}
\usepackage{graphicx}
\usepackage{placeins}
\usepackage{dsfont}
\usepackage{subcaption}
\usepackage{bbm}
\usepackage{natbib}
\usepackage{url} % not crucial - just used below for the URL 
\usepackage{xcolor}
\usepackage{multirow}
\usepackage[normalem]{ulem}
\newcommand{\blind}{0}

\newcommand{\gscale}{\tau}
\newcommand{\lscale}{\lambda}
\newcommand{\Lscale}{\Lambda}
\newcommand{\X}{X}
\newcommand{\y}{y}
\newcommand{\coef}{\beta}
\newcommand{\s}{\sigma}
\newcommand{\transpose}{\text{\raisebox{.5ex}{$\intercal$}}}
\newcommand{\diff}{\operatorname{\mathrm{d}}\!{}}
\newcommand{\eqDistribution}{\mathrel{\raisebox{-.2ex}{$\overset{\scalebox{.6}{$\, d$}}{=}$}}}
\newcommand\numberthis{\addtocounter{equation}{1}\tag{\theequation}}
\newcommand{\piglo}{\pi_\mathrm{glo}}
\newcommand{\piloc}{\pi_\mathrm{loc}}
\newcommand{\cdf}{\textsc{cdf}}
\newcommand{\loggscale}{\gscale_\mathrm{log}} % You can adjust the notation as you see fit.
\newcommand{\unnormInd}{^\dagger}
\newcommand{\ess}{\textsc{ess}}
\newcommand{\indicator}{\mathds{1}}
\newcommand{\halfspace}{\mskip.5\thinmuskip}
\newcommand{\given}{\mskip.85\thinmuskip | \mskip.85\thinmuskip}

\definecolor{emerald}{RGB}{12, 166, 151}
\definecolor{lava}{rgb}{0.81, 0.06, 0.13}

\begin{document}

\def\spacingset#1{\renewcommand{\baselinestretch}%
	{#1}\small\normalsize} \spacingset{1}

%%%%%%%%%%%%%%%%%%%%%%%%%%%%%%%%%%%%%%%%%%%%%%%%%%%%%%%%%%%%%%%%%%%%%%%%%%%%%%

\if0\blind
{
	\title{\bf Spectral Collapsed Gibbs Sampler for Bayesian Sparse Regression}
	\author{Andrew Chin
		%\thanks{The authors gratefully acknowledge \textit{please remember to list all relevant funding sources in the unblinded version}}\hspace{.2cm}
		\\
		%Department of Biostatistics, Johns Hopkins University\\
		and \\
		Xiyu Ding \\
		%Department of Biostatistics, Johns Hopkins University\\
		and \\
		Akihiko Nishimura\\
		Department of Biostatistics, Johns Hopkins University\\}
	\maketitle
} \fi

\if1\blind
{
	\bigskip
	\bigskip
	\bigskip
	\begin{center}
		{\LARGE\bf  Spectral Collapsed Gibbs Sampler for Bayesian Sparse Regression}
	\end{center}
	\medskip
} \fi

\bigskip
\begin{abstract}
Sparse regression based on global-local shrinkage priors are increasingly used for Bayesian modeling of modern high-dimensional data, but scaling up the Gibbs sampler for posterior inference remains a challenge. 
While much effort has gone into speeding up the high-dimensional coefficient update step, insufficient attention has been given to the potential poor mixing of the global scale parameter $\gscale$ and of the overall sampler.
One proposed remedy has been to marginalize out the coefficients when updating $\gscale$.
Here we show that, while this collapsed update was previously thought to require a Metropolis step, we can in fact sample directly and efficiently from the collapsed density.
This is made possible by careful linear algebraic manipulations and a strategic per-Gibbs-scan spectral decomposition, allowing subsequent evaluations of the collapsed density across hundreds of values of $\gscale$ at negligible cost.
We combine this computational trick with adaptive numerical integration and inverse transform sampling to construct a direct sampler.
This eliminates the need to tune Metropolis proposals and yields faster convergence and improved mixing.
We demonstrate our method on two big data applications, fitting logistic regression under the horseshoe prior to datasets with design matrices of size $120{,}000 \times 1{,}379$ and $1{,}980 \times 17{,}848$.
\end{abstract}

\noindent%
\vfill

\newpage
\spacingset{1.75} % DON'T change the spacing!
\section{Introduction}
\label{sec:intro}
Continuous shrinkage priors play an important role in modern Bayesian inference, allowing for sparsity to be induced in posteriors.
In this article, we consider the popular global-local priors, such as the horseshoe \citep{carvalho2009handling} and bridge \citep{polson2014bayesian}, and their use in  linear and logistic models. 
Such sparse regression models have found a variety of applications, including in genetics, proteomics, and disease forecasting \citep{lee2020estimation, vanarsa2023comprehensive, zhang2022gene}.
Global-local shrinkage priors on the regression coefficients $\beta_j$ are parametrized as
\begin{equation}\label{eq:linregprior}
	\beta_j \mid \lambda, \gscale^2 \sim \operatorname{N}(0, \gscale^2 \lambda^2_j), \quad \gscale \sim \piglo(\gscale) , \quad \lscale_j \sim \piloc(\lscale_j).
\end{equation}
The global scale parameter $\gscale \in \mathbb{R^+}$ determines the amount of overall sparsity in the coefficients, and the local scale parameters $\lscale_j \in \mathbb{R^+}$ allow specific coefficients to be estimated away from zero.
Different choices of $\piloc$ give rise to different classes of global-local priors.

\begin{figure}
	\begin{subfigure}{.5\textwidth}
		\centering
		\includegraphics[width=\linewidth]{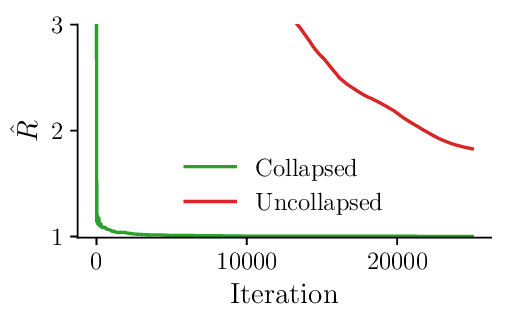}
	\end{subfigure}%
	\begin{subfigure}{.5\textwidth}
		\centering
		\includegraphics[width=\linewidth]{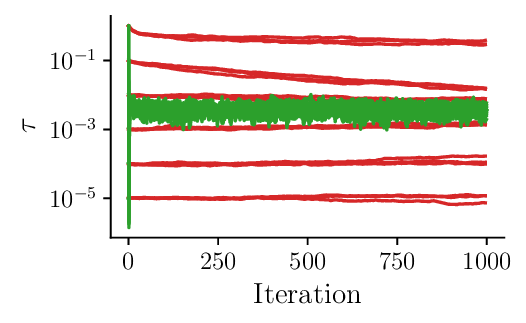}
	\end{subfigure}
	\caption{%
	Comparison of the collapsed and uncollapsed Gibbs samplers' convergence behaviors in the \textsc{ehr} data application of Section~\ref{subsec:logregpn};
	details of the experimental set-ups are provided in Section~\ref{sec:app}, but we display part of the results here to illustrate the issue of poor mixing in the uncollapsed sampler.
	Both samplers draw directly, without any use of a Metropolis step and the like, from the collapsed and uncollapsed densities when updating $\gscale$.
	The right figure shows trace plots of $\gscale$, from six different initializations, for the first 1000 iterations.
	The left figure shows how the rank-normalized $\hat R$ for $\gscale$ evolves over iterations.
	A commonly used threshold for declaring convergence is 1.01, which the uncollapsed sampler does not come close to approaching.
	}
	\label{fig:collapse_full_convergence}
\end{figure}

Posterior inference under these models is typically carried out through Gibbs sampling \citep{bhattacharya2022geometric, johndrow2020scalable}.
The most obvious computational bottleneck is the update of the high-dimensional regression coefficients $\beta$, and much effort has gone into improving its efficiency  \citep{bhattacharya2016fast, hahn2019efficient, nishimura2023prior}.
Another bottleneck, which has received far less attention despite its critical performance impact, is the poor mixing in the global scale~$\gscale$.
The standard full conditional update of $\gscale$ is straightforward computationally, but its conditioning on the high-dimensional $\beta$ can lead to poor mixing in the resulting Markov chain.
While this issue has not been widely discussed in the literature, it has been pointed out by \cite{polson2014bayesian} and is also evident in our real-data examples (Figure~\ref{fig:collapse_full_convergence}).
To help remedy this, \citet{johndrow2020scalable} propose marginalizing out $\beta$ in the update of $\gscale$ and sampling from this collapsed density via the Metropolis algorithm.
This collapsing significantly improves mixing and, while evaluating the collapsed density is now computationally more expensive, often delivers substantial improvements in overall computational efficiency.

In this work, we develop computational techniques that enhance the efficiency and practicality of the collapsing approach.
Speeding up the collapsed update of $\gscale$ is important since it constitutes the computational bottleneck for the entire Gibbs sampler.
To this end, we start by observing that the requisite computations for evaluating the collapsed density can be categorized into three groups based on their required frequencies:
(1) once per dataset prior to the Gibbs sampling, (2) once per Gibbs scan, with their results reusable for different values of $\gscale$, and (3) once per each value of $\gscale$, even within the same Gibbs scan.
Then, having observed the majority of the requisite computations to fall into Categories (1) and (2), we identify a previously overlooked opportunity to reduce the cost of Category (3) computations from cubic to linear in $\min\{n, p\}$, the smaller of the sample size or the number of predictors.

We achieve the cost reduction in Category~(3) computations through careful linear algebraic manipulations and a strategic once-per-Gibbs-scan application of a spectral decomposition.
This makes it possible to evaluate, within each Gibbs scan, the collapsed density at many values of $\gscale$ at negligible additional cost.
We take advantage of this computational trick to obtain a high-fidelity approximation of the cumulative density function (\cdf)  through adaptive numerical integration, which in turn allow us to sample directly from the density via inverse transform sampling  \citep{devr86, derflinger2010random}.
We thus eliminate the reliance on the Metropolis update for $\gscale$, which not only removes the need to tune its proposal variance but also enhances the sampler's convergence and mixing.

We demonstrate our method on two logistic regression models.
The first uses plasmode synthetic genotype data from \citet{DVN/COXHAP_2022}, with $n = 120{,}000$ individuals and $p = 1{,}379$ genetic markers, to model single nucleotide polymorphisms' (\textsc{snp}) effects on a simulated phenotype.
The second is a propensity score estimation model for two second-line type-2 diabetes treatments, and uses $n = 1{,}980$ patient records on $p = 17{,}848$ indicators of clinical history from Johns Hopkins electronic health records (\textsc{ehr}).

\section{Gibbs sampling sparse regression posteriors}
\label{sec:methods}

We begin by discussing the Gibbs samplers for global-local priors and their updates for $\gscale$.
First, we derive the collapsed update which marginalizes out $\beta$ for both linear and logistic models.
Then, we look at strategies for sampling from these collapsed densities.
In addition to a standard Metropolis sampler, we highlight the potential use of numerical integration for this purpose, which is often not discussed within the Bayesian computation literature but can be suitable for single-dimensional densities.

\subsection{Collapsed Gibbs sampler for linear regression}
\label{subsec:model}
First we consider a linear model for a continuous outcome $y$ with design matrix $X \in \mathbb{R}^{n\times p}$.
We focus on the conjugate normal-gamma formulation 
\begin{equation}\label{eq:y}
	\y\sim \operatorname{N}(\s\X\coef, \s^2I_n),\quad  \s^2 \sim \operatorname{InvGamma}(a, b),
\end{equation}
where the inclusion of $\s$ in the mean allows us to collapse $\s$ in Gibbs updates.

A standard Gibbs sampler updates $\gscale$ from its full conditional $\gscale \mid \lscale, \beta, \s^2, \X, \y$, direct samplers for which are available under common choices of a prior $\pi(\gscale)$.
The collapsed Gibbs sampler of \citet{johndrow2020scalable} marginalizes out both $\coef$ and $\s^2$ from the update of $\gscale$ to improve mixing and cycles through the following conditional updates---to simplify notation, we henceforth omit $\y$ and $\X$ from the conditioning:
\[
 1. \ \gscale \mid \lscale \,; \quad 
 2. \  \s^2 \mid \lscale,\gscale \,; \quad 
 3. \ \coef \mid \s^2 , \lscale, \gscale \,; \quad 
 4. \  \lscale \mid \gscale, \coef, \s^2.
\]
Step 2 and 3 are conjugate updates, involving inverse gamma and multivariate Gaussian distributions, respectively.
For Step 4, efficient rejection samplers are available \citep{johndrow2020scalable, nishimura2023shrinkage}.
When opting for the full conditional update, the high-dimensional update of the coefficients in Step~3 constitutes the main computational bottleneck.
The collapsed Gibbs sampler has roughly the same overall computational cost as the uncollapsed sampler, but the previously most expensive quantities are now first computed in Step~1 and later reused in Step~3.
For the collapsed sampler, therefore, we can treat Step~1 as the main computational bottlneck for the entire Gibbs sampler.

Our focus is on Step 1, which traditionally relied on a Metropolis step.
Since the likelihood $\y \mid \lscale, \gscale,\s^2 \sim N(X\beta \s, \s^2)$ and prior $\beta_j \mid \gscale, \lscale, \s^2 \sim N(0, \gscale^2 \lscale_j^2)$ are both Gaussian, it is straightforward to marginalize out $\beta$ from the likelihood to obtain
\begin{equation}
\label{eq:partially_collaped_linear_likelihood}
\y \mid \lscale, \gscale,\s^2 \sim N(0, \s^2 M_\gscale)
	\ \text{ where } \,
	M_\gscale = I_n + \gscale^2X\Lscale^2X^\transpose 
	\, \text{ and } \, 
	\Lscale = \textrm{diag}(\lscale).
\end{equation}
Under the conjugate prior $\s^2 \sim \operatorname{InvGamma}(a,b)$, we can further marginalize out $\s^2$ to obtain the marginal likelihood
\begin{align*}
L(\y \mid \lambda, \tau) 
	%&= \int\int \pi(\beta, \s^2, y \mid \lscale, \gscale) \diff{\beta}\diff\s^2\\
	%&\propto \int\int \pi(\beta, \s^2, y, \lscale, \gscale) \diff{\beta}\diff\s^2\\	
	%&\propto \int\int \pi(\y \mid \beta, \s^2, \lscale, \gscale)\pi(\beta)\pi(\s^2) \diff{\beta}\diff\s^2\\
	%&\propto \int\pi(\s^2) \int \underbrace{\pi(\y \mid \beta, \s^2)}_{N(\X\beta\s, \sigma^2)}\underbrace{\pi(\beta)}_{N(0, \tau^2\Lscale^2)} \diff{\beta}\diff\s^2\\
	%&\propto \int \underbrace{\pi(\s^2)}_{InvGamma(a,b)}  \ \underbrace{\pi(\y \mid \lscale, \gscale, \s^2 )}_{N(0, \s^2M_\gscale)}  \diff\s^2\\
	&\propto \left|M_\gscale\right|^{-1/2}
	\left( b+\frac{1}{2}\y^\transpose  M_\gscale^{-1}\y \right)^{-\left( \frac{n}{2}+a \right)},
\end{align*}
\noindent where $| \cdot |$ denotes the determinant.
Therefore, the collapsed Gibbs updates $\gscale$ from
\begin{equation}\label{eq:condpost}
\pi(\gscale \mid  \lscale) \propto  
	\left|M_\gscale\right|^{-1/2}
	\left( b+\frac{1}{2}y^\transpose  M_\gscale^{-1}y \right)^{-\left(\frac{n}{2}+a \right)}
	\pi(\gscale).
\end{equation}

\subsection{Collapsing in logistic regression}
\label{subsec:logreg}
Here we show how to adapt our collapsing approach to the logistic regression with likelihood
\[
y_i \mid \beta \sim \operatorname{Binomial}(n_i, p_i)
	\ \text{ for } \, 
	\operatorname{logit}(p_i) = x_i^\transpose \beta.
\]
\noindent We employ the P\'{o}lya--Gamma augmentation of \citet{polson2013bayesian}.
Conditionally on the auxiliary P\'{o}lya--Gamma variable $\omega \in \mathbb{R^+}^p$, the logistic likelihood is transformed into a form
$$z_i \mid \beta, \omega \sim \operatorname{N}(x_i^\transpose \beta, \omega_i^{-1})
	\ \text{ for } \,
	z_i = y_i - n_i/2,$$
the Gaussianity of which allows us to integrate out $\beta$.
% and $\beta \mid \omega, \Lambda, \tau \sim \operatorname{N}(X^\transpose \Omega X + \tau^{-2}\Lambda^{-2})$.
Unlike in the linear case where the variance parameter $\s^2$ can be additionally integrated out, here we cannot integrate out $\omega$.
The collapsed sampler thus updates $\gscale$ from the density
\begin{equation}\label{eq:logreg_cond_posterior}
\begin{aligned}
\pi(\gscale \mid \lscale, \omega) 
	&\propto  
	\left|M_\gscale\right|^{-1/2}
	\exp\!\left( \frac{1}{2}z^\transpose  M_\gscale^{-1} z \right)
	\pi(\gscale)\\
&\hspace*{-1em}\text{ where } \,
	M_\gscale = \Omega^{-1} + \gscale^2X\Lscale^2 X^\transpose
	\, \text{ and } \, 
	\Omega = \textrm{diag}(\omega).
\end{aligned}
\end{equation}
Putting this all together and noting that conditional independence implies $\lscale \mid \gscale, \coef, \omega \eqDistribution \lscale \mid \gscale$ and $\omega \mid  \lscale, \gscale, \beta \eqDistribution \omega \mid \beta$, the collapsed Gibbs sampler for logistic regression cycles through the following conditional updates:
\begin{equation}\label{eq:logreg_gibbs}
1. \ \gscale \mid \lscale, \omega \,; \quad 2. \ \coef \mid \lscale,  \gscale, \omega \,; \quad 3. \ \lscale \mid \gscale, \coef \,; \quad 4. \  \omega \mid \beta.
\end{equation}

\subsection{Sampling the collapsed density: Metropolis vs.\ direct sampling via high-fidelity numerical inverse transform}
\label{subsec:int}
We now consider two potential approaches to dealing with the collapsed conditionals $\pi(\gscale \mid \lscale)$ and $\pi(\gscale \mid \lscale, \omega)$ of Equations~\eqref{eq:condpost} in \eqref{eq:logreg_cond_posterior}, corresponding to the linear and logistic cases, both of which lie outside standard parametric families.
Our discussion here is agnostic to an actual form of the density, so we refer to both collapsed densities generically as $\pi(\gscale \given \ast \halfspace)$ and their unnormalized versions as $\pi\unnormInd(\gscale \given \ast \halfspace)$.

The simplest way to deal with $\pi(\gscale \given \ast \halfspace)$ is to resort to the Metropolis algorithm, as done in \cite{johndrow2020scalable} for the linear model.
The collapsed update then would require two (unnormalized) density evaluations, one at the current value and the other at the proposal value.
If the cost of evaluating $\pi\unnormInd(\gscale \given \ast \halfspace)$ at each $\gscale$ value was a tiny fraction of the overall cost, however, we could instead consider numerically integrating the density to obtain its \cdf{} and draw directly using inverse transform sampling \citep{devr86, derflinger2010random}.
This approach not only obviates the need to tune Metropolis proposals, but also offers an opportunity to significantly improve mixing.

\begin{figure}[tb]
	\begin{center}
		\includegraphics[width=\textwidth]{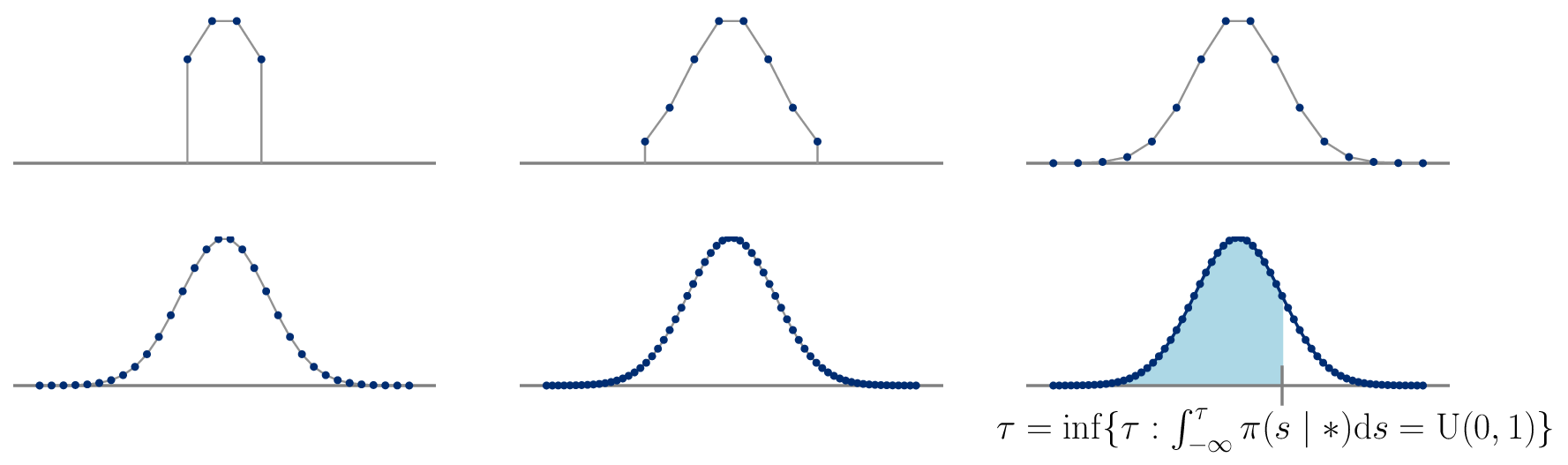}
	\end{center}
	\vspace{-3mm}
	\caption{%
	Illustration of the adaptive numerical integration scheme to construct a high-fidelity approximation of the \cdf{} for inverse transform sampling.
	First, from a rough estimate of the mode, a grid is expanded outwards to find reasonable upper and lower bounds for integration (top row). 
	The approximation is then refined by halving the grid spacing until the change in integral, evaluated via trapezoid rule, becomes negligible between two successive approximations (bottom row). 
	Finally, the resulting numerically integrated \cdf{} is used to inverse transform a uniform random variable (bottom right).
	}
	\label{fig:numer_int}
\end{figure}

Deferring to Section~\ref{sec:efficient_computation} the question of how evaluating $\pi\unnormInd(\gscale \given \ast \halfspace)$ can be made so inexpensive, we now provide practical details on how to implement the direct sampling approach.
The key step is generating a high-fidelity \cdf{} approximation via adaptive numerical integration.
We target the log-transformed parameter $\loggscale = \log \gscale$ because we find the density of $\loggscale$ to vary more smoothly and be more amenable to numerical integration.
The goal then is to approximate
$\Pi(\loggscale \mid \ast) = \int_{-\infty}^{\loggscale} \pi\unnormInd(\loggscale' \mid \ast) \diff \loggscale' \, /  \int_{-\infty}^\infty \pi\unnormInd( \loggscale' \mid \ast) \diff \loggscale'$ by numerically integrating the unnormalized density $\pi\unnormInd$ over a fine grid of points.
Our algorithm constructs the grid as follows.

We first need to identify a range of numerical integration so that it covers essentially all the probability mass of $\pi(\gscale \given \ast \halfspace)$.
To this end, we take the value of $\loggscale$ from a previous cycle of the Gibbs sampler and, from there, expand a grid of size $\Delta \loggscale = 1$  in both directions until the relative change in the trapezoidal approximation of the total integral $\int_0^\infty \pi( \loggscale' \mid \ast) \diff \loggscale'$ is less than 0.1\% and the endpoint densities are less than $0.0001$.
Having identified the integration range, we refine the approximation by repeatedly halving the grid spacing until two successive approximations achieve the relative change in integral of less than 0.1\%.
With this high-fidelity approximation in hand, we can apply the inverse \cdf{} to a $\operatorname{Unif}(0,1)$ variable and generate a sample from $\pi(\gscale \given \ast \halfspace)$,
as illustrated in Figure \ref{fig:numer_int}.

We find the above basic adaptive method with trapezoidal rule works well in all our applications, but more advanced quadrature algorithms \citep{press2007numerical} can also be considered.
Pseudo code for the adaptive integration procedure is provided in Supplement~\ref{supp:integrator}.

\section{%
	Fast simultaneous evaluations of the collapsed density at multiple $\gscale$ values}
\label{sec:efficient_computation}

We have so far derived the formulas for the collapsed densities and laid out the algorithmic framework to deal with the collapsed update, either through a Metropolis or the direct sampler via high-fidelity numerical inverse transform.
Now we discuss how to efficiently carry out the collapsed density evaluations and what operations constitute the computational bottlenecks.
To this end, we start by observing that requisite computations to evaluate $\pi\unnormInd(\gscale \given \ast \halfspace)$ fall into three categories based on their required frequencies of evaluation: 

\begin{enumerate}
	\item One evaluation per dataset: Computation of quantities, like $X^\transpose X$ and $X^\transpose y$, that are independent of unknown parameters. 
	Since these can be precomputed before Gibbs sampling, we treat the cost of computing these as negligible.
	\item One evaluation per Gibbs scan: Computations of quantities, like $X \Lscale^2 X^\transpose$ and $X^\transpose \Omega X$, that are independent of $\gscale$.
	\item One evaluation per value of $\gscale$: Computations of quantities, like $|M_\gscale|$ and $y^\transpose  M_\gscale^{-1}y$ for $M_\gscale$ as defined in Equation~\eqref{eq:partially_collaped_linear_likelihood} and \eqref{eq:logreg_cond_posterior}, that depend on $\gscale$ and require additional computations beyond those in Category~2 and 3, even within the same Gibbs scan.
\end{enumerate}

We show in Section~\ref{subsec:spectral_collapsing_p_bigger} and \ref{subsec:spectral_collapsing_n_bigger} that, by exploiting suitable linear algebraic identities and applying the spectral decomposition to an appropriate square matrix of size $\min\{n, p\}$ within per-scan computations, we can make the relative cost of subsequent per-$\gscale$ computations  essentially negligible.
In turn, this allows us to directly sample from the collapsed density through the numerical inverse transform method of Section~\ref{subsec:int}.
In Section~\ref{subsec:eig_vs_chol}, we compare the computational complexity of our spectral collapsed update to that of the Metropolis update.
We show that, while the Metropolis update only requires two Cholesky decompositions and is technically faster computationally, the additional cost for the direct update is generally small enough to be offset by the improved mixing. 
Table~\ref{tab:chol_vs_spectral} summarizes the main computational costs and relative merits of our spectral collapsing approach.

\begin{table}[tb]
	\caption{%
		Computational complexity, as derived in Section~\ref{subsec:eig_vs_chol}, of evaluating $\pi\unnormInd(\gscale \given \ast \halfspace)$ with the Cholesky and spectral decomposition-based approaches.
		The latter is more expensive for the first evaluation but renders subsequent evaluations negligible in cost. 
		The table shows the $p>n$ regime for brevity; the $n>p$ regime is given by interchanging $n$ and $p$.}
	\label{tab:chol_vs_spectral}
	\vspace{-0.5\baselineskip}
	\spacingset{1.55}
	\begin{center}
		{\setlength{\tabcolsep}{11pt}
			\begin{tabular}{c|cc}
				&  \multicolumn{2}{c}{Complexity ($p > n$)}\\
				& Cholesky & Spectral \\
				\hline
				First evaluation& $2n^2p+ \frac{2}{3}n^3$ & $2n^2p + \frac{8}{3}n^3$ \\			
				Subsequent evaluation & $2n^2p+ \frac{2}{3}n^3$ & $n$ \\
			\end{tabular}
		}
	\end{center}
	\spacingset{1.75}
	\vspace{-.5\baselineskip}
\end{table}

\subsection{Spectral collapsing for fast per-$\gscale$ evaluation: $\boldsymbol{p>n}$ regime}
\label{subsec:spectral_collapsing_p_bigger}
In devising the most efficient approach to evaluating the collapsed density, it is important to treat the regimes $p > n$ and $n > p$ differently.
We first consider the $p>n$ regime, where $\mathcal{O}(n^2p)$ and $\mathcal{O}(n^3)$ operations are more cost-effective than $\mathcal{O}(np^2)$ and $\mathcal{O}(p^3)$ ones.
For the linear case, recall from Equation~\eqref{eq:condpost} that the collapsed density takes the form
\begin{equation}
\label{eq:pn_M}
	\pi(\gscale \mid  \lscale) \propto  \left|M_\gscale\right|^{-1/2}\left(b+\frac{1}{2}y^\transpose  M_\gscale^{-1}y\right)^{-\left(\frac{n}{2}+a\right)}\pi(\gscale)
	\ \, \text{ for } \,
	M_\gscale = I_n + \gscale^2X\Lscale^2X^\transpose.
%	\text{ and } 
%	\Lscale = \operatorname{diag}(\lambda).
\end{equation}
In evaluating the above, the most computationally intensive operations are: forming the matrix product $X\Lscale^2X^\transpose$ as part of computing $M_\gscale$, calculating the determinant $| M_\gscale |$, and evaluating the quadratic form $y^\transpose M_\gscale^{-1}y$ involving the matrix inverse.

The matrix product $X\Lscale^2X^\transpose$ is a per-scan $\mathcal{O}(n^2p)$ operation and, in the $p > n$ regime, is the most expensive one.
The subsequent evaluations of $|M_\gscale|$ and $y^\transpose M_\gscale^{-1}y$ can be carried out via a per-$\gscale$ $\mathcal{O}(n^3)$ Cholesky decomposition, which is the standard approach to calculating a determinant and solving a positive definite system.
This per-$\gscale$ $\mathcal{O}(n^3)$ cost is less expensive than the per-scan $\mathcal{O}(n^2p)$ cost, but unless $p \gg n$, can quickly add up to become a bottleneck if evaluating the collapsed density at hundreds of $\gscale$ values for the direct sampling method of Section~\ref{subsec:int}.

To make the direct sampling possible, we now present an alternative approach that renders the cost of subsequent per‑$\tau$ computations $\mathcal{O}(n)$ and hence essentially negligible.
%utilizes linear algebraic identities and a single \textit{per-scan} $\mathcal{O}(n^3)$ spectral decomposition to
The idea is to apply a single per-scan $\mathcal{O}(n^3)$ spectral decomposition 
\[X\Lscale^2X^\transpose = VD_nV^\transpose,\] 
where $V$ is the orthogonal matrix of eigenvectors and $D_n = \operatorname{diag}(\{ d_i \}_{i=1}^n)$ is the diagonal matrix of eigenvalues; 
importantly, both $V$ and $D_n$ are independent of $\gscale$.
Computing the determinant is now $\mathcal{O}(n)$ since adding the identity matrix only shifts the eigenvalues by 1:
\begin{equation}\label{eq:det}
	\begin{split}
		\left|M_\gscale\right| 
		= \prod_{i=1}^n \operatorname{eig}(M_\gscale)_i 
		= \prod_{i=1}^n \operatorname{eig} \! \left( I_n + \gscale^2 VD_nV^\transpose \right)_i 
		%&= \prod_{i=1}^n \{1 + \gscale^2 \operatorname{eig}( VDV^\transpose)_i \} \\
		=  \prod_{i=1}^n \left( 1+\gscale^2d_{i} \right).
	\end{split}
\end{equation}
The spectral decomposition also allows us to reduce the cost of computing $y^\transpose  M_\gscale^{-1}y$ to $\mathcal{O}(n)$ through the following linear algebraic identity: 
\begin{align*}
	M_\gscale^{-1} 
	&= (I_n + \gscale^2 VD_nV^\transpose)^{-1}\\
%	&= I_n - \gscale^2 V(D^{-1}+ \gscale^2V^\transpose V)^{-1}V^\transpose\\
	&= I_n - \gscale^2 V(D_n^{-1}+ \gscale^2I_n)^{-1}V^\transpose\\
	&= I_n - \gscale^2 V\textrm{diag}\left(\left\{\frac{1}{d_i^{-1}+\gscale^2}\right\}_i\right)V^\transpose,
\end{align*}
which allows us to compute the quadratic form as
\begin{align*}
	y^\transpose  M_\gscale^{-1}y &= y^\transpose\left\{   I_n - \gscale^2 V\operatorname{diag}\left(\frac{1}{d_i^{-1}+\gscale^2}\right)V^\transpose \right\}y\\
	%&= y^\transpose y - \tau^2 y^\transpose V \textrm{diag}\left(\frac{1}{d_i^{-1}+\tau^2}\right) V^\transpose y \\
	&=  \underbrace{y^\transpose y \vphantom{\frac{1}{1_1^{-1}}}}_{\mathcal{O}(n) \text{ per-dataset}} + 
	\underbrace{y^\transpose  V\vphantom{\frac{1}{1_1^{-1}}}}_{\mathcal{O}(n^2) \text{ per-scan}}
	\underbrace{\textrm{diag}\left(\left\{\frac{\gscale^{2} }{d_i^{-1}+\gscale^{2}}\right\}\right)}_{\mathcal{O}(n) \text{ per-$\gscale$}}
	\underbrace{V^\transpose y\vphantom{\frac{1}{1_1^{-1}}}}_{\mathcal{O}(n^2)\text{ per-scan}}.  \numberthis \label{eq:quadratic}
	% there has to be a better way to tex the \vphantom  but it works for now
\end{align*} 
Since the vector $y^\transpose V$ remains constant across $\gscale$ and the multiplication by the diagonal matrix %$\textrm{diag}\left(\left\{ \gscale^{2} / (d_i^{-1}+\gscale^{2}) \right\}\right)$ 
is $\mathcal{O}(n)$, the total cost of evaluation for each additional $\gscale$ is also $\mathcal{O}(n)$.

Our spectral collapsing technique is also applicable to the collapsed Gibbs sampler \eqref{eq:logreg_gibbs} for logistic regression with the auxiliary P\'{o}lya--Gamma parameter $\omega$.
There, $M_\gscale$ is given as in Equation~\eqref{eq:logreg_cond_posterior} and can be expressed as 
\begin{equation}\label{eq:logreg_pn_M}
M_\gscale = \Omega^{-1/2} (I_n  + \gscale^2\Omega^{1/2}X\Lscale^2 X^\transpose\Omega^{1/2}) \Omega^{-1/2}.
\end{equation}
A spectral decomposition $\Omega^{1/2}X\Lscale^2 X^\transpose\Omega^{1/2} = VD_nV^\transpose$ then allows us to quickly evaluate the density~\eqref{eq:logreg_cond_posterior} using the following identities, analogous to the linear case:
\begin{equation}\label{eq:logreg_pn_det}
	\begin{split}
		\left|M_\gscale\right| &= \left|\Omega^{-1/2}\right|\left|I_n  + \gscale^2VD_nV^\transpose\right|\left|\Omega^{-1/2}\right| 
		= \left(\prod_i \omega_i^{-1}\right)\prod_i \left(1+\gscale^2 d_i \right) 
	\end{split}
\end{equation}
and
\begin{equation}\label{eq:logreg_pn_quad}
z^\transpose M_\gscale^{-1} z = z^\transpose \Omega z -  z^\transpose \Omega^{1/2} V \halfspace\operatorname{diag}\!\left(\frac{\gscale^2}{d_i^{-1}+\gscale^2}\right)V^\transpose \Omega^{1/2} z.
\end{equation}

Equipped with the fast per-$\gscale$ evaluation of the densities \eqref{eq:condpost} and \eqref{eq:logreg_cond_posterior} as above, each $\gscale$ update of our spectral collapsed Gibbs sampler can be summarized as follows:
\begin{enumerate}
	\item Compute the spectral decomposition $\X \Lscale \X^\transpose = VD_nV^\transpose$ (linear case) or $\Omega^{1/2}X\Lscale^2 X^\transpose\Omega^{1/2} \linebreak = VD_nV^\transpose$ (logistic case) to enable the $\mathcal{O}(n)$ evaluation of the density for each $\gscale$.
	\item Use the adaptive numerical integration scheme of Section~\ref{subsec:int} to construct the high-fidelity approximation $\hat F(\loggscale)$ to the \cdf{} of the log-transformed parameter $\loggscale$.
	\item Draw $u\sim  \operatorname{Unif}(0,1)$, set $\loggscale = F^{-1}(u)$, and return $\gscale = \exp(\loggscale)$.
\end{enumerate}

\subsection{Spectral collapsing in $\boldsymbol{n>p}$ regime}
\label{subsec:spectral_collapsing_n_bigger}

We have observed the key bottlenecks in evaluating $\pi\unnormInd(\gscale \given \ast \halfspace)$ to be computing $M_\gscale$, calculating the determinant $| M_\gscale |$, and evaluating the quadratic form $y^\transpose M_\gscale^{-1}y$.
We have also observed the main per-scan computational costs of spectral collapsing in the $p > n$ regime to be
the $\mathcal{O}(n^2p)$ matrix product $X\Lscale^2X^\transpose$ (linear case) or $\Omega^{1/2}X\Lscale^2 X^\transpose\Omega^{1/2}$ (logistic case), 
and the $\mathcal{O}(n^3)$ spectral decomposition of the resulting $n \times n$ matrix.
In the $n > p$ regime, through an application of the Woodbury and other linear algebraic identities, we reduce the main per-scan computational costs to a $\mathcal{O}(n p^2)$ matrix product and a $\mathcal{O}(p^3)$ decomposition, and the subsequent per-$\gscale$ to $\mathcal{O}(p)$.

Starting from the linear case, we observe that
\begin{equation}\label{eq:np_woodbury}
	M_\gscale^{-1} = (I_n + \gscale^2X\Lscale^2X^\transpose)^{-1} = I_n -  \gscale^2 X \Lscale (I_p + \gscale^2 \Lscale X^\transpose X\Lscale )^{-1} \Lscale  X^\transpose
\end{equation}
and carry out a $p \times p$ spectral decomposition 
\begin{equation}
\label{eq:spec_decomp_for_np_case}
\Lambda X^\transpose  X \Lambda = V D_p V^\transpose,
\end{equation}
where $D_p = \operatorname{diag}(\{ d_i \}_{i=1}^p)$.
The determinant $| M_\gscale |$ can now be evaluated at $\mathcal{O}(p)$ because
\begin{align*}
	\left|I_n + \gscale^2 X \Lambda^2 X^\transpose\right| 
		&= \left|\gscale^{-2}\Lambda^{-2} + X^\transpose X\right| \left|I_p^{-1}\right|\left|\gscale^2  \Lambda^2\right| 
		= \left|I_p + \gscale^2\Lambda X^\transpose X\Lambda \right|
		= \prod_{i = 1}^p \left(1+\gscale^2 d_i \right),
\end{align*}
where the first equality follows from the identity $|A + U B W^\transpose| = |B^{-1} + W^\transpose A^{-1} U| |B|  |A|$ for $W, U \in \mathbb{R}^{n \times p} $ and invertible $A \in \mathbb{R}^{n \times n}, B \in \mathbb{R}^{p\times p}$ \citep{harville1998matrix}.
The quadratic form $y^\transpose M_\gscale^{-1} y $ also admits an $\mathcal{O}(p)$ evaluation via 
\begin{equation}\label{eq:linreg_np_quad}
	\begin{split}
		y^\transpose M_\gscale^{-1} y 
		&= y^\transpose \left[ I_n - \gscale^2  X \Lambda \left\{ I_p + \gscale^2 VD_pV^\transpose \right\}^{-1} \Lambda X^\transpose \right] y \\
		&= y^\transpose  y - y^\transpose  X \Lscale V \operatorname{diag}\!\left(\frac{\gscale^2}{1+\gscale^2 d_i}\right) V^\transpose \Lscale X^\transpose y,
	\end{split}
\end{equation}
where the latter equality follows from the fact 
\begin{equation}\label{eq:inverse}
\left\{ I_p + \gscale^2 VD_pV^\transpose \right\}^{-1} 
= I_p - \gscale^2 V \operatorname{diag} \!\left(\frac{1}{d_i^{-1}+\gscale^2}\right) V^\transpose
= V \operatorname{diag}\left( \frac{1}{1+\gscale^2 d_i} \right) V^\transpose.	
\end{equation}

For the logistic case, we can achieve $\mathcal{O}(p)$ evaluations for the determinant and the quadratic form in an analogous manner.
We first observe 
\begin{equation}\label{eq:logreg_np_woodbury}
\begin{split}
M_\gscale^{-1} &= (\Omega^{-1} + \gscale^2 X \Lambda^2 X^\transpose)^{-1} \\
&= \Omega - \Omega X ( \gscale^{-2}\Lambda^{-2} +  X^\transpose \Omega X)^{-1} X^\transpose \Omega \\
&= \Omega - \gscale^2\Omega X \Lambda(I_p + \gscale^2\Lambda X^\transpose \Omega X \Lambda)^{-1}\Lambda X^\transpose \Omega
\end{split}
\end{equation}
and carry out a $p \times p$ spectral decomposition $\Lambda X^\transpose \Omega X \Lambda = VD_pV^\transpose$.
The $\mathcal{O}(p)$ evaluations are now possible through the identities
\begin{equation}\label{eq:logreg_np_det}
	\begin{split}
		\left| \Omega^{-1} + \gscale^2 X \Lambda^2 X^\transpose \right|
		= \left|\gscale^{-2}\Lambda^{-2} + X^\transpose \Omega X\right| \left|\Omega^{-1}\right|\left|\gscale^2  \Lambda^2\right| 
%		&= |\gscale^{-2}\Lambda^{-2}| |I + \gscale^2\Lambda X^\transpose \Omega X\Lambda | |\Omega^{-1}||\gscale^2  \Lambda^2| \\
		&= \left(\prod_i \omega_i^{-1}\right)\prod_i \left(1+\gscale^2 d_i \right) 
	\end{split}
\end{equation}
and
\begin{equation}\label{eq:logreg_np_quad}
	\begin{split}
z^\transpose M_\gscale^{-1} z 
&= z^\transpose \left[ 
		\Omega - \gscale^2 \Omega X \Lambda \left\{ I_p + \gscale^2 VD_pV^\transpose  \right\}^{-1} \Lambda X^\transpose \Omega
	\right] z \\
%&= z^\transpose \left[\Omega - \gscale^2 \Omega X \Lambda \left\{I - \gscale^2 V\textrm{diag}\left(\frac{1}{d_i^{-1}+\gscale^2}\right)V^\transpose\right\} \Lambda X^\transpose \Omega \right] z \\
&= z^\transpose \Omega z + 
	z^\transpose \Omega X \Lscale V \operatorname{diag}\!\left(\frac{\gscale^2}{1+\gscale^2 d_i}\right) V^\transpose \Lscale X^\transpose\Omega z.
\end{split}
\end{equation}

Having derived the fast per-$\gscale$ evaluation in both $p > n$ and $n > p$ regimes, we close the discussion by describing how our spectral collapsing can additionally accommodate estimating a subset of coefficients without shrinkage.
For example, an intercept is usually given a Gaussian prior $\beta_0 \sim N(0, \sigma_0^2)$, separately from the global-local shrinkage on the other coefficients \citep{tadesse2021handbook_bayes_variable_select}.
In this case, the joint coefficient vector $\beta \given \gscale, \lscale$ has a prior 
%$\beta \given \gscale, \lscale \sim N\left(0, \operatorname{diag}( \{ \sigma_0^2, \gscale \lscale_1^1, \ldots, \gscale^2 \lscale_p\} ) \right)$
\[
\beta \given \gscale, \lscale, \sigma_0 \sim N\left(0, \begin{bmatrix} \sigma_0^2 & 0\\ 0 & \gscale^2 \Lscale^2 \end{bmatrix}\right),
\]
which prevents us from pulling the factor $\gscale^2$ out to leave the rest of the covariance independent of the global scale.
Nonetheless, the fast per-$\gscale$ evaluation can still be achieved with appropriate modifications.
Specifically, we handle the relevant computations blockwise when $n>p$ or through the Woodbury formula when $p>n$, with minimal additional per-scan and per-$\gscale$ costs.
Further details are provided in Supplement~\ref{supp:intercept}.

\subsection{Cost comparison of spectral- and Cholesky-based approaches}
\label{subsec:eig_vs_chol}

\begin{figure}[h]
\vspace{-1.5mm}
	\begin{center}
		\includegraphics[width=0.6\textwidth]{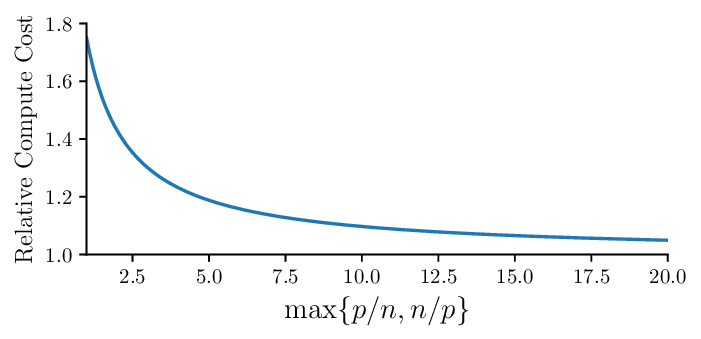}
	\end{center}
	\vspace{-8mm}
	\caption{%
	Relative costs of the spectral and Metropolis updates of $\gscale$ as a function of $p / n$ in the $p > n$ regime (and of $n / p$ in the $n > p$ regime).
	The costs in the $p > n$ regime are $2n^2p + 8n^3/3$ and $2n^2p + 2n^3/3$ for the spectral and Metropolis updates, respectively;
	the costs in the $n > p$ regime are given by switching the roles of $n$ and $p$.
	For given $n$ and $p$, the direct sampling via the spectral update has an advantage provided that the gain in per-iteration mixing  exceeds the increase in the relative cost---%
	this in particular holds true in the example applications of Section~\ref{sec:app}, where we observe two- to eightfold overall efficiency improvements over optimally and suboptimally tuned Metropolis samplers.
	%up to threefold overall efficiency improvements over optimally tuned Metropolis samplers and up to eightfold over suboptimally tuned versions.
	}
	\label{fig:compute_compare}
\end{figure}

As mentioned in the introduction to Section~\ref{sec:efficient_computation}, compared to the Cholesky-based Metropolis update of $\gscale$, the use of the spectral decomposition in our approach incurs additional cost.
%The spectral collapsed update of $\gscale$ has computational advantage  when the additional cost is more than offset by improved mixing.
%This additional cost is beneficial depends on how much mixing benefit is gained from directly sampling $\gscale$.
We now analyze its contribution to the overall computational cost as $n$ and $p$ vary.
The update via spectral collapsing has an advantage whenever improvement in mixing outweighs the additional cost.
We conduct complexity analysis of each approach, using the number of floating point operations as a measure of each algorithm's cost.

We first focus on the $p>n$ regime.
Both spectral and Metropolis updates of $\gscale$ require $2n^2p$ operations for the matrix product $X\Lscale^2X^\transpose$ as part of forming the $n \times n$ matrix $M_\gscale$.
Additionally, the spectral update requires $8n^3/3$ operations for a spectral decomposition of $M_\gscale$, while the Metropolis update requires $2n^3/3$ operations for two Cholesky decompositions \citep{trefethen2022numerical}.
In total, the spectral and Metropolis updates require $2n^2p + 8n^3/3$ and $2n^2p + 2n^3/3$, respectively.
Figure~\ref{fig:compute_compare} compares the relative cost between the two updates---%
it illustrates how the additional cost of spectral collapsing is relatively small, especially as the difference in scale between $n$ and $p$ grows.
Our numerical experiments of Section~\ref{sec:app} also find that the improvements in mixing more than pay for the additional cost.
%Even when the additional cost is non-negligible, we find in Section~\ref{sec:app} that the improvements in mixing can easily make up for it.

In the $n>p$ regime, the costs are given by switching the roles of $p$ and $n$;
i.e.\ $2p^2 n + 8p^3/3$ and $2p^2 n + 2p^3/3$ for the spectral and Metropolis updates, respectively.
This result applies only to the logistic case, however.
For the linear case in the $n>p$ regime, the expensive matrix product required for density evaluation is $X^\transpose X$ (Equation~\eqref{eq:spec_decomp_for_np_case}), which can be computed once per dataset and reused during the Gibbs sampling.
This means that the requisite per-scan computations for the spectral and Metropolis updates are bottlenecked only by the spectral and Cholesky decompositions.
This results in overall costs of $8p^3/3$ and $2p^3/3$ for the two updates
%, without the additional $2p^2 n$ term, 
and in a higher relative cost for the spectral update. 
Even in this case, direct sampling of the spectral update retains the advantage of being tuning-free, while the mixing of the Metropolis update can be sensitive to its tuning.
%Note that while $X\Lscale^2X^\transpose$ still appears in Equation~\eqref{eq:linreg_np_quad}, it is as the quadratic form $y^\transpose X \Lscale^2 X^\transpose y$ which can be computed via $\mathcal{O}(pn)$ matrix-vector operations.

\section{Applications}\label{sec:app}
We now demonstrate our spectral collapsed Gibbs sampler on two large-scale logistic regression models, one in the $n>p$ regime and the other in the $p>n$ regime.
In both applications, we use the horseshoe prior on regression coefficients, with $\piloc \sim \textrm{HalfCauchy}(0,1)$, and assign $\gscale$ a $\operatorname{Unif}(0,1)$ prior.

We compare three update strategies for $\gscale$: 
the uncollapsed full conditional update, the Metropolis collapsed update, and the spectral collapsed update via inverse transform sampling.
The uncollapsed density can be represented as a gamma distribution, truncated to $[0, 1]$ by the prior, and can be drawn directly. 
For the Metropolis collapsed update, we consider a Gaussian random-walk proposal with three alternative choices of the variance:
one is chosen to achieve an acceptance rate of 44\%, which is optimal for one dimensional Gaussian targets and is often recommended for univariate targets \citep{gelman1996efficient, gelman2013bda}, 
and the others are chosen by multiplying and dividing this value by four to assess the resulting Gibbs sampler's sensitivity to tuning.
%The spectral collapsed update uses the direct sampling approach as detailed in Section~\ref{subsec:int}.
%In total, we have five Gibbs samplers that differ only in their updates of $\gscale$ and update the rest of the parameters in the same manner:
%$\lscale$ is updated via the rejection sampler from Appendix F of \cite{nishimura2023shrinkage}, while the rest are updated directly from the respective parametric families.
With the three alternative tunings for the Metropolis collapsed update, we in total have five Gibbs samplers that differ only in their update schemes for $\gscale$.
These Gibbs samplers otherwise employ the same update strategies for the rest of the parameters:
$\lscale$ is updated via the rejection sampler from Appendix F of \cite{nishimura2023shrinkage} and the rest directly from their respective parametric families.

For all the Gibbs samplers, we initialize $\gscale$ at $10^{-k}$ for $k = 0, \ldots, 5$, covering a sufficient range of values to investigate the samplers' behaviors when started from under- and over-estimated initial values of $\gscale$.
For each initial value of $\gscale$, we run three independent chains. 
We initialize $\lscale$ as values drawn from the prior and $\omega$ as $1/2$, its conditional mean given $\beta = 0$.
The uncollapsed sampler additionally requires an initial value for $\beta$, which we draw from its distribution conditional on the other parameters' initial values.

We compare the samplers in terms of both rate of convergence and mixing efficiency.
We measure the former by the number of iterations until the rank-normalized $\hat R$ for $\gscale$ falls below 1.01 across all the 18 independent chains, obtained from the 6 initial $\gscale$ values  and 3 seeds \citep{vehtari2021rank, arviz_2019}.
The latter is measured by the effective sample size (\ess{}) of the post-convergence samples \citep{plummer2006coda} normalized by computational time.
We run all the samplers on identical hardware to ensure fair comparisons in terms of computational time.
Code for the samplers is available at \url{https://tinyurl.com/bkaxahce}.

\subsection{$\boldsymbol{n>p}$ logistic regression for genetic risk prediction}
\label{subsec:logregnp}

Our first application considers the problem of predicting individuals' risks of developing a disease based on their genetic profiles \citep{chatterjee2016developing_prs}.
For this purpose, we use the synthetic genotype dataset from \citet{DVN/COXHAP_2022}, filtered to HapMap3 \textsc{snp}s \citep{oliver_pain_2023_7773502}, with $120{,}000$ subjects and $1{,}177{,}528$ indicators of \textsc{snp}s.
To simulate a realistic binary disease phenotype $y$, we first draw 0.1\% of the \textsc{snp} coefficients $\beta$ from a standard normal variable and set the rest to 0, and generate a latent phenotype as $\y' \sim \operatorname{N}(\X \beta, 40^2)$.
This yields a signal to noise ratio, defined as $\textrm{var}(X\beta)/\textrm{var}(y')$, of approximately 0.2 to align with a degree of genetic heritability seen in the real world \citep{dun2024robust}.
We then set $y = \indicator(y' > c)$ with $c=60$ so that the binary outcome roughly has a 10\% prevalence.
Finally, we apply clumping \citep{prive2019clumping}, which takes advantage of the correlation among \textsc{snp}s to select a subset that retains most of the overall predictive power, and obtain $p=1{,}379$ indicators to be regressed on the phenotype $y$ of $n=120{,}000$ subjects.

\begin{table}[tb]
	\caption{Convergence and mixing efficiency for each sampler in the $n>p$ example.
	Note the \ess{} estimates, which were computed using the last 5,000 iterations, may not be reliable for chains which have not converged. 
	The metrics are chosen so that smaller values indicate greater efficiency.}
	\label{tab:logreg_resultsnp}
	\vspace{-0.5\baselineskip}
	\spacingset{1.55}
	\begin{center}
		\begin{tabular}{c|ccc}
			& Iterations until & Iterations & Minutes  \\
			Method  	& $\hat R < 1.01$ & per \ess{} &  per \ess{} \\
			\hline
			Metropolis Collapsed (optimal)& 5140 & 20 & 7.62\\			
			Metropolis Collapsed (optimal$\times4$) & 4575 & 45  & 18.04	\\
			Metropolis Collapsed (optimal$/4$)& $>10000$ & 55  & 22.15	\\
			Spectral Collapsed & \textbf{490} & \textbf{7}  & \textbf{2.67}	\\
			Uncollapsed & 2285 & 11  & 3.93	\\
			% used 5000 (50%) for burn in 
		\end{tabular}
	\end{center}
	\spacingset{1.75}
	\vspace{-.5\baselineskip}
\end{table}

\begin{figure}[htb]
	\begin{subfigure}{.5\textwidth}
		\centering
		\includegraphics[width=.95\linewidth]{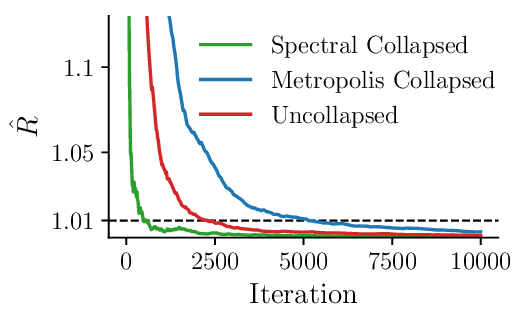}
	\end{subfigure}%
	\begin{subfigure}{.5\textwidth}
		\centering
		\includegraphics[width=.95\linewidth]{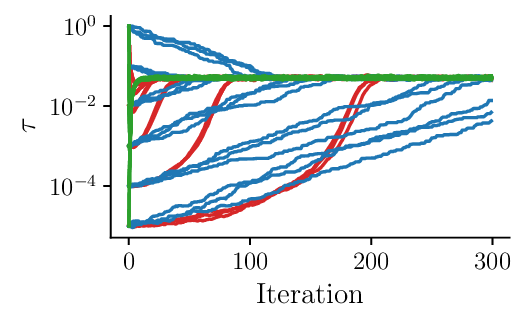}
	\end{subfigure}
	\caption{%
		Evolution of the convergence diagnostic $\hat{R}$ (left) and trace plots of $\gscale$ from different initializations (right).
		The spectral collapsed sampler reaches $\hat R<1.01$ and converges to a high density region in fewer iterations than the optimally tuned Metropolis collapsed sampler. 
		The two collapsed samplers have comparable per-iteration costs in this example, with the spectral sampler costing no more than 5\% additional computing time. 
		The spectral sampler also outperforms the uncollapsed sampler, but the latter outperforms the Metropolis sampler in this $n > p$ example, where the benefit of collapsing turns out relatively small; see Figure~\ref{fig:compare_conditionals} and its caption for more detailed discussions.
		To avoid cluttering the plot, the results for the non-optimal Metropolis samplers are not shown.}
	\label{fig:snp_convergence}
\end{figure}

All the Gibbs samplers are run for 10,000 iterations.
As shown in Table~\ref{tab:logreg_resultsnp} and Figure~\ref{fig:snp_convergence}, we observe that the spectral collapsed sampler almost immediately finds the high density region and converges ten times faster than the optimal Metropolis sampler.
It also shows superior post-convergence mixing, delivering threefold improvement in \ess{}.
Here, the magnitude of improvement remains approximately the same whether normalizing by iterations or by time;
this is because the spectral decomposition adds negligible cost to the Cholesky given the relative sizes of $n$ and $p$ in this example.
The Metropolis sampler's performance further degrades when using suboptimal proposal variances.

\begin{figure}[h]
	\begin{center}
		\includegraphics[width=.95\textwidth]{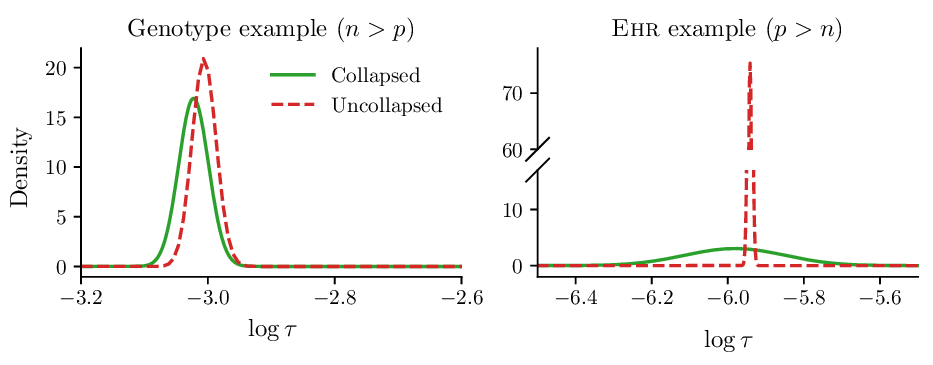}
	\end{center}
	\vspace*{-.75\baselineskip}
	\caption{%
		Comparison of the collapsed and uncollapsed densities when conditioning on the same values for the remaining parameters, taken from the last iteration  of a run of the spectral collapsed Gibbs sampler for illustration. 
		The relationship between the two densities remain similar when conditioning on other values.
		In the $n>p$ genotype data example, the uncollapsed density differ only slightly from the collapsed one, making the uncollapsed Gibbs sampler reasonably effective. 
		In the $p>n$ \textsc{ehr} data example, however, the uncollapsed density grossly underestimates the posterior uncertainty in $\gscale$, causing the entire chain to mix poorly.
		Analogous behavior of an uncollapsed Gibbs sampler is observed in the context of Gaussian process models, where collapsing is critical for efficient updates of hyperparameters and for overall mixing \citep{filippone2014pseudo_marginal}.
		%the conditional distributions of hyperparameters must also be marginalized in order to successfully sample posteriors.
		}
	\label{fig:compare_conditionals}
\end{figure}

We also find that, in this example, the uncollapsed sampler outperforms the Metropolis collapsed sampler.
This is due to the fact that the uncollapsed density here happens to differ only slightly from the collapsed one (Figure~\ref{fig:compare_conditionals}); 
the mixing benefit of direct sampling thus outweighs that of collapsed sampling. 
The spectral sampler combines the benefits of both direct and collapsed samplings, and outperforms the uncollapsed sampler by about 30\%. 
The uncollapsed sampler's solid performance here should not be viewed as representative;
the next $p>n$ example will demonstrate how the full conditional update can fail catastrophically.

\subsection{$\boldsymbol{p>n}$ logistic regression in an observational study using {\small EHR}}
\label{subsec:logregpn}

Our second example considers the problem of estimating the propensity score using the large-scale approach of \citet{tian2018large_scale_ps}, for the purpose of estimating the causal effect of assignment to two alternative treatments.
That is, we fit a high-dimensional logistic regression with the treatment assignment indicator as the outcome.
We use a dataset of type-2 diabetes patients, extracted from \textsc{ehr} at the Johns Hopkins Health System according to the new-user cohort design as specified in the protocol of \citet{khera2023legendt2dm}.
These patients are assigned to one of the two second-line treatments, dipeptidyl peptidase-4 inhibitors and glucagon-like peptide-1 receptor agonists.
As potential predictors of the treatment assignment, we extract the patients' demographics and clinical histories, including: medication usage, conditions, clinical observations, diagnoses, and lab measurements. 
This results in a design matrix representing $n=1{,}980$ patients and $p=17{,}848$ covariates.

We run both collapsed samplers for 25,000 iterations.
We run the uncollapsed sampler for at least the same amount of computing time as the spectral collapsed sampler;
since the uncollapsed update is faster in this example, this results in the uncollapsed sampler run for 150,000 iterations.

\begin{table}[htb]
	\caption{Convergence and mixing efficiency for each sampler in the $p>n$ example, with \ess{} computed using the last 10,000 iterations.
	As in Table~\ref{tab:logreg_resultsnp}, metrics are chosen so that smaller values indicate greater efficiency.}

	\label{tab:logreg_resultspn}
	\vspace{-0.5\baselineskip}
	\spacingset{1.55}
	\begin{center}
		\begin{tabular}{c|ccc}
			& Iterations until & Iterations & Minutes  \\
		Method  	& $\hat R < 1.01$ & per \ess{} &  per \ess{} \\
		\hline
		Metropolis Collapsed (optimal) & 10835 & 45 & 1.32\\
		Metropolis Collapsed (optimal$\times4$) & 11510 & 87  & 2.50\\
		Metropolis Collapsed (optimal$/4$)& $24535$ & 121 & 3.67\\
		Spectral Collapsed & \textbf{5350} & \textbf{27} & \textbf{1.13} \\
		Uncollapsed & $>150000$ & 756 & 69.35	\\
		% used 12000 for burn in
		\end{tabular}
	\end{center}
	\spacingset{1.75}
	\vspace{-.5\baselineskip}
\end{table}

\begin{figure}[htb]
	\begin{subfigure}{.5\textwidth}
		\centering
		\includegraphics[width=\linewidth]{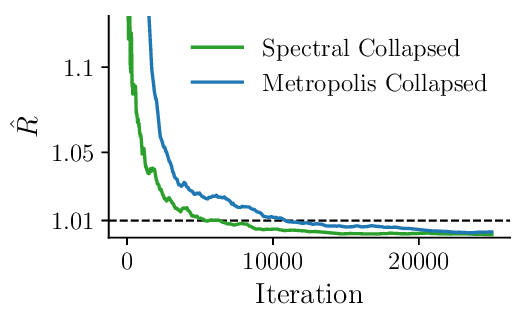}
	\end{subfigure}%
	\begin{subfigure}{.5\textwidth}
		\centering
		\includegraphics[width=\linewidth]{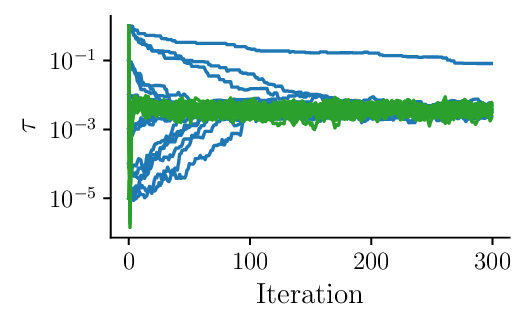}
	\end{subfigure}
	\caption{
		As in Figure~\ref{fig:snp_convergence}, the spectral collapsed method finds the high density region nearly immediately, and convergence happens twice as quickly as the optimally tuned Metropolis sampler.
		The uncollapsed sampler converges so slowly that, in order to avoid it obscuring the difference between the two collapsed sampler, we plot it separately in Figure~\ref{fig:collapse_full_convergence}.
	}\label{fig:jhu_convergence}
\end{figure}

As in the example of Section~\ref{subsec:logregnp}, the spectral collapsed method outperforms the Metropolis collapsed update by achieving faster convergence and better mixing (Table \ref{tab:logreg_resultspn}). 
Since the difference in the magnitudes of $n$ and $p$ are smaller than in the previous example, the cost of the spectral decomposition is more noticeable here, with the spectral sampler taking roughly 1.3 times longer per iteration than the Metropolis sampler.
This results in comparable performance between the two collapsed samplers when accounting for computation time.
However, our spectral collapsing method has a significant practical advantage of being tuning-free.
In contrast, the Metropolis update's performance is sensitive to the choice of proposal variance, the tuning of which in practice must be done adaptively during a burn-in period and may take a substantial and unknown number of iterations before the sampler reaches its near-optimal performance \citep{andrieu2008adaptive_mcmc}.

\begin{figure}
	\begin{center}
		\includegraphics[width=\textwidth]{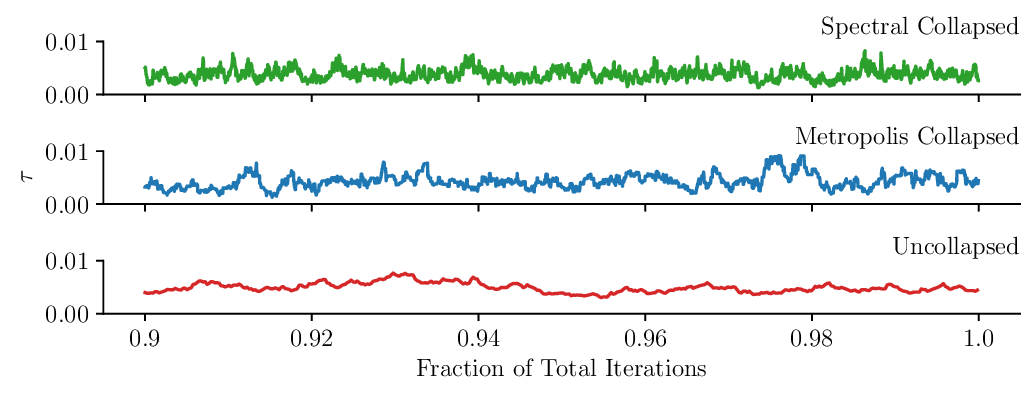}
	\end{center}
	\vspace{-0.7\baselineskip}
	\caption{
	Trace plots of the global scale over the last 10\% of iterations of each sampler.
	We show the outputs from the samplers initialized from $\gscale = 10^{-3}$, the value closest to stationarity, since the uncollapsed sampler otherwise fails to converge.
	The spectral collapsed sampler exhibits the best mixing, while the uncollapsed sampler with full conditional update struggles.}\label{fig:logreg_logp}
\end{figure}

The uncollapsed sampler faces a major mixing issue in this $p > n$ example, never reaching $\hat R<1.01$ within the allocated time (Table~\ref{tab:logreg_resultspn} and Figure~\ref{fig:collapse_full_convergence}).
The trace plot in Figure~\ref{fig:logreg_logp} further illustrates the severity of this issue.
The poor convergence and mixing here indicate that, in the $p > n$ regime with large $p$, collapsing over the high-dimensional regression coefficients is critical in ensuring efficient exploration of the global scale parameter $\gscale$. 
While we are not the first to point this out \citep{polson2014bayesian,johndrow2020scalable}, the issue remains under-explored in the literature and merits further investigation to more precisely characterize regimes in which the collapsed update becomes essential.

\section{Discussion}\label{sec:discussion}
In this article, we have devised a computational technique that leverages the spectral decomposition to enable fast multi-point evaluations of the collapsed density of $\gscale$ and direct sampling from it.
We demonstrate, using the two real-world applications to cover both $n > p$ and $p > n$ regimes, how the spectral collapsing method offers significant practical advantages in posterior computation for Bayesian sparse regression.
The numerical results also reinforce the importance of the collapsed update in high-dimensional problems.
More generally, an uncollapsed update has been found to be problematic when updating a model hyperparameter that depends on a large number of other parameters \citep{filippone2014pseudo_marginal}.
Such bottlenecks have received comparatively less attention in the computational statistics literature, and addressing them constitutes an important area of future research.

%
%\bigskip
%\begin{center}
%{\large\bf SUPPLEMENTARY MATERIAL}
%\end{center}
%
%\begin{description}
%
%\item[Python Code:] Python code for all the simulations run in the article is available at \todo{github....}
%
%\item[Data:] Data for the applications is available at TODO
%
%\end{description}

\bibliographystyle{apalike}
\spacingset{1.15} 
\bibliography{refs}

@article{arviz_2019,
  doi = {10.21105/joss.01143},
  url = {https://doi.org/10.21105/joss.01143},
  year = {2019},
  publisher = {The Open Journal},
  volume = {4},
  number = {33},
  pages = {1143},
  author = {Ravin Kumar and Colin Carroll and Ari Hartikainen and Osvaldo Martin},
  title = {ArviZ a unified library for exploratory analysis of {B}ayesian models in {P}ython},
  journal = {Journal of Open Source Software}
}

@article{andrieu2008adaptive_mcmc,
  title={A tutorial on adaptive {MCMC}},
  author={Andrieu, Christophe and Thoms, Johannes},
  journal={Statistics and Computing},
  volume={18},
  number={4},
  pages={343--373},
  year={2008},
  publisher={Springer}
}

@article{bhattacharya2016fast,
  title={Fast sampling with {G}aussian scale mixture priors in high-dimensional regression},
  author={Bhattacharya, Anirban and Chakraborty, Antik and Mallick, Bani K},
  journal={Biometrika},
  volume={103},
  number={4},
  pages={985--991},
  year={2016},
  publisher={Oxford University Press}
}

@article{bhattacharya2022geometric,
  title={Geometric ergodicity of {G}ibbs samplers for the Horseshoe and its regularized variants},
  author={Bhattacharya, Suman and Khare, Kshitij and Pal, Subhadip},
  journal={Electronic Journal of Statistics},
  volume={16},
  number={1},
  pages={1--57},
  year={2022},
  publisher={The Institute of Mathematical Statistics and the Bernoulli Society}
}

@inproceedings{carvalho2009handling,
  title={Handling sparsity via the horseshoe},
  author={Carvalho, Carlos M and Polson, Nicholas G and Scott, James G},
  booktitle={Artificial Intelligence and Statistics},
  pages={73--80},
  year={2009},
  organization={PMLR}
}

@article{chatterjee2016developing_prs,
  title={Developing and evaluating polygenic risk prediction models for stratified disease prevention},
  author={Chatterjee, Nilanjan and Shi, Jianxin and Garc{\'\i}a-Closas, Montserrat},
  journal={Nature Reviews Genetics},
  volume={17},
  number={7},
  pages={392--406},
  year={2016},
  publisher={Nature Publishing Group}
}

@article{derflinger2010random,
  title={Random variate generation by numerical inversion when only the density is known},
  author={Derflinger, Gerhard and H{\"o}rmann, Wolfgang and Leydold, Josef},
  journal={ACM Transactions on Modeling and Computer Simulation (TOMACS)},
  volume={20},
  number={4},
  pages={1--25},
  year={2010},
  publisher={ACM New York, NY, USA}
}

@Book{devr86,
  Title                    = {Non-Uniform Random Variate Generation},
  Author                   = {Luc Devroye},
  Publisher                = {Springer-Verlag},
  Year                     = {1986},
  Address                  = {New York, NY, USA}
}

@article{dun2024robust,
  title={A Robust {B}ayesian Method for Building Polygenic Risk Scores using Projected Summary Statistics and Bridge Prior},
  author={Dun, Yuzheng and Chatterjee, Nilanjan and Jin, Jin and Nishimura, Akihiko},
  journal={arXiv preprint arXiv:2401.15014},
  year={2024}
}

@article{filippone2014pseudo_marginal,
  title={Pseudo-marginal {B}ayesian inference for {G}aussian processes},
  author={Filippone, Maurizio and Girolami, Mark},
  journal={IEEE Transactions on Pattern Analysis and Machine Intelligence},
  volume={36},
  number={11},
  pages={2214--2226},
  year={2014},
  publisher={IEEE}
}

@article{gelman1996efficient,
  title={Efficient {M}etropolis jumping rules},
  author={Gelman, Andrew and Roberts, Gareth O and Gilks, Walter R},
  journal={Bayesian Statistics},
  volume={5},
  pages={599--608},
  year={1996},
  publisher={Oxford University PressOxford}
}

@book{gelman2013bda,
  title     = {Bayesian Data Analysis},
  edition   = {3},
  author    = {Gelman, Andrew and Carlin, John B. and Stern, Hal S. and Dunson, David B. and Vehtari, Aki and Rubin, Donald B.},
  year      = {2013},
  publisher = {Chapman and Hall/CRC},
  address   = {Boca Raton, FL},
  isbn      = {9781439840955}
}

@article{hahn2019efficient,
  title={Efficient sampling for {G}aussian linear regression with arbitrary priors},
  author={Hahn, P Richard and He, Jingyu and Lopes, Hedibert F},
  journal={Journal of Computational and Graphical Statistics},
  volume={28},
  number={1},
  pages={142--154},
  year={2019},
  publisher={Taylor \& Francis}
}

@book{harville1998matrix,
  author = {David A. Harville},
  title = {Matrix Algebra From a Statistician's Perspective},
  publisher={Springer-Verlag},
  address= {New York, NY, USA},
  year={1997},
}

@article{johndrow2020scalable,
  author  = {James Johndrow and Paulo Orenstein and Anirban Bhattacharya},
  title   = {Scalable Approximate {MCMC} Algorithms for the Horseshoe Prior},
  journal = {Journal of Machine Learning Research},
  year    = {2020},
  volume  = {21},
  number  = {73},
  pages   = {1--61},
  url     = {http://jmlr.org/papers/v21/19-536.html}
}

@article{khera2023legendt2dm,
  title   = {Multinational patterns of second line antihyperglycaemic drug initiation across cardiovascular risk groups: federated pharmacoepidemiological evaluation in {LEGEND-T2DM}},
  author  = {Khera, Rohan and Dhingra, Lovedeep Singh and Aminorroaya, Arya and Li, Kelly and Zhou, Jin J. and Arshad, Faaizah and Blacketer, Clair and Bowring, Mary G. and Bu, Fan and Cook, Michael and Dorr, David A. and Duarte-Salles, Talita and DuVall, Scott L. and Falconer, Thomas and French, Tina E. and Hanchrow, Elizabeth E. and Horban, Scott and Lau, Wallis C. Y. and Li, Jing and Liu, Yuntian and Lu, Yuan and Man, Kenneth K. C. and Matheny, Michael E. and Mathioudakis, Nestoras and McLemore, Michael F. and Minty, Evan and Morales, Daniel R. and Nagy, Paul and Nishimura, Akihiko and Ostropolets, Anna and Pistillo, Andrea and Posada, Jose D. and Pratt, Nicole and Reyes, Carlen and Ross, Joseph S. and Seager, Sarah and Shah, Nigam and Simon, Katherine and Wan, Eric Y. F. and Yang, Jianxiao and Yin, Can and You, Seng Chan and Schuemie, Martijn J. and Ryan, Patrick B. and Hripcsak, George and Krumholz, Harlan and Suchard, Marc A.},
  journal = {BMJ Medicine},
  year    = {2023},
  volume  = {2},
  number  = {1},
  pages   = {e000651},
  month   = oct,
  doi     = {10.1136/bmjmed-2023-000651},
  url     = {https://bmjmedicine.bmj.com/content/2/1/e000651}
}

@article{lee2020estimation,
  title={Estimation of {COVID}-19 spread curves integrating global data and borrowing information},
  author={Lee, Se Yoon and Lei, Bowen and Mallick, Bani},
  journal={PloS One},
  volume={15},
  number={7},
  pages={e0236860},
  year={2020},
  publisher={Public Library of Science San Francisco, CA USA}
}

@article{nishimura2023prior,
  title={Prior-preconditioned conjugate gradient method for accelerated Gibbs sampling in “large n, large p” Bayesian sparse regression},
  author={Nishimura, Akihiko and Suchard, Marc A},
  journal={Journal of the American Statistical Association},
  volume={118},
  number={544},
  pages={2468--2481},
  year={2023},
  publisher={Taylor \& Francis}
}

@article{nishimura2023shrinkage,
  title={Shrinkage with shrunken shoulders: {G}ibbs sampling shrinkage model posteriors with guaranteed convergence rates},
  author={Nishimura, Akihiko and Suchard, Marc A},
  journal={Bayesian Analysis},
  volume={18},
  number={2},
  pages={367--390},
  year={2023},
  publisher={International Society for Bayesian Analysis}
}

@dataset{oliver_pain_2023_7773502,
  author       = {Oliver Pain},
  title        = {HapMap3 SNP-list},
  month        = mar,
  year         = 2023,
  publisher    = {Zenodo},
  doi          = {10.5281/zenodo.7773502},
  url          = {https://doi.org/10.5281/zenodo.7773502},
}

@article{plummer2006coda,
  title={{CODA}: convergence diagnosis and output analysis for {MCMC}},
  author={Plummer, Martyn and Best, Nicky and Cowles, Kate and Vines, Karen and others},
  journal={R News},
  volume={6},
  number={1},
  pages={7--11},
  year={2006}
}

@article{polson2013bayesian,
  title={Bayesian inference for logistic models using {P}{\'o}lya--Gamma latent variables},
  author={Polson, Nicholas G and Scott, James G and Windle, Jesse},
  journal={Journal of the American Statistical Association},
  volume={108},
  number={504},
  pages={1339--1349},
  year={2013},
  publisher={Taylor \& Francis}
}

@article{polson2014bayesian,
  title={The bayesian bridge},
  author={Polson, Nicholas G and Scott, James G and Windle, Jesse},
  journal={Journal of the Royal Statistical Society: Series B: Statistical Methodology},
  pages={713--733},
  year={2014},
  publisher={JSTOR}
}

@article{prive2019clumping,
  title        = {Making the Most of Clumping and Thresholding for Polygenic Scores},
  author       = {Priv{\'e}, Florian and Vilhj{\'a}lmsson, Bjarni J. and Aschard, Hugues and Blum, Michael G. B.},
  journal      = {The American Journal of Human Genetics},
  year         = {2019},
  volume       = {105},
  number       = {6},
  pages        = {1213--1221},
  month        = dec,
  publisher    = {Elsevier},
  doi          = {10.1016/j.ajhg.2019.11.001},
  issn         = {0002-9297}
}

@book{press2007numerical,
  title={Numerical recipes 3rd edition: The art of scientific computing},
  author={Press, William H and Teukolsky, Saul A and Vetterling, William T and Flannery, Brian P},
  year={2007},
  publisher={Cambridge university press}
}

@article{tadesse2021handbook_bayes_variable_select,
  title={Handbook of {B}ayesian variable selection},
  author={Tadesse, Mahlet G and Vannucci, Marina},
  year={2021},
  publisher={CRC Press}
}

@article{tian2018large_scale_ps,
  title={Evaluating large-scale propensity score performance through real-world and synthetic data experiments},
  author={Tian, Yuxi and Schuemie, Martijn J and Suchard, Marc A},
  journal={International Journal of Epidemiology},
  volume={47},
  number={6},
  pages={2005--2014},
  year={2018},
  publisher={Oxford University Press}
}

@book{trefethen2022numerical,
  title={Numerical Linear Algebra},
  author={Trefethen, L.N. and Bau, D.},
  isbn={9780898719574},
  series={Other Titles in Applied Mathematics},
  year={1997},
  publisher={Society for Industrial and Applied Mathematics}
}

@article{vanarsa2023comprehensive,
  title={Comprehensive proteomics and platform validation of urinary biomarkers for bladder cancer diagnosis and staging},
  author={Vanarsa, Kamala and Castillo, Jessica and Wang, Long and Lee, Kyung Hyun and Pedroza, Claudia and Lotan, Yair and Mohan, Chandra},
  journal={BMC Medicine},
  volume={21},
  number={1},
  pages={1--17},
  year={2023},
  publisher={BioMed Central}
}

@article{vehtari2021rank,
author = {Aki Vehtari and Andrew Gelman and Daniel Simpson and Bob Carpenter and Paul-Christian B{\"u}rkner},
title = {{Rank-Normalization, Folding, and Localization: An Improved $\widehat{R}$ for Assessing Convergence of {MCMC} (with Discussion)}},
volume = {16},
journal = {Bayesian Analysis},
number = {2},
publisher = {International Society for Bayesian Analysis},
pages = {667 -- 718},
year = {2021},
doi = {10.1214/20-BA1221},
URL = {https://doi.org/10.1214/20-BA1221}
}

@article{zhang2022gene,
  title={Gene-environment interactions explain a substantial portion of variability of common neuropsychiatric disorders},
  author={Zhang, Hanxin and Khan, Atif and Rzhetsky, Andrey},
  journal={Cell Reports Medicine},
  volume={3},
  number={9},
  year={2022},
  publisher={Elsevier}
}

@data{DVN/COXHAP_2022,
author = {Zhang, Haoyu},
publisher = {Harvard Dataverse},
title = {{Simulated data for 600,000 subjects from five ancestries}},
year = {2022},
version = {V5},
doi = {10.7910/DVN/COXHAP},
url = {https://doi.org/10.7910/DVN/COXHAP}
}
\spacingset{1.75} 
\newpage

\appendix
\section{Adaptive numerical integrator for high-fidelity approximation of {\large CDF}}
\label{supp:integrator}

\newcommand{\grid}{\mathpzc{T}}
\newcommand{\gridElem}[1]{\gscale_{\mathrm{log}, \halfspace #1}}
\newcommand{\integChangeTol}{\epsilon_\mathrm{integral}}
\newcommand{\densityChangeTol}{\epsilon_\mathrm{density}}
\newcommand{\densit}{\mathcal{F}}
\newcommand{\density}{f}

\spacingset{1.55} 
\begin{algorithm}
	\caption{Adaptive Numerical Integrator}\label{alg:int}
	\begin{algorithmic}[1]
		\Require Initial value $\gridElem{1} $, unnormalized density $\pi\unnormInd$, integral change threshold $\integChangeTol$, endpoint density threshold $\densityChangeTol$, grid expansion size $\Delta$.
		%\State $x \gets[x-\delta, x', x+\delta]$
		%\State $f \gets [\log f(x_0), \dots, \log f(x_n)] $
		%\State $z_{\textrm{new}} \gets \max(f)$ \Comment{Log shift constant}
		%\State $F_{\textrm{new}} \gets Trapezoid(f-z)$ \Comment{Integral value}
		\State $(F_{\textrm{old}}, F_{\textrm{new}}) \gets (0, 1)$ \Comment{Placeholder values for the first iteration of the while loop}
		\State $\grid \gets [\gridElem{1}]$ \Comment{Evaluation grid}
		\State ${f} = [f_1, \ldots, f_{| \grid |} ] \gets \pi\unnormInd( \grid)$ \Comment{Density values, evaluated pointwise}
		\While{$(F_{\textrm{new}} - F_{\textrm{old}})/F_{\textrm{old}} > \integChangeTol$ or ${f}_{1}/\max f > \densityChangeTol$ or ${f}_{| \grid |}/\max f> \densityChangeTol$}
		\State $F_{\textrm{old}} \gets F_{\textrm{new}}$
		\State $\grid \gets [ \gridElem{1} -\Delta, \grid,  \gridElem{| \grid |}+\Delta]$  \Comment{Expand grid on both ends by $\Delta$}  
		\State ${f} \gets \pi\unnormInd(\grid)$
		\State $F_{\textrm{new}} \gets \text{Trapezoid}(\grid, f)$
		\EndWhile
		\While{$(F_{\textrm{new}} - F_{\textrm{old}})/F_{\textrm{old}} > \integChangeTol$} 
		\State $F_{\textrm{old}} \gets F_{\textrm{new}}$
		\State $\displaystyle \grid \gets \left[ \gridElem{1}, \frac{ \gridElem{1} + \gridElem{2} }{2},  \gridElem{2}, \frac{ \gridElem{2} +  \gridElem{3} }{2}, \dots, \frac{  \gridElem{ | \grid | - 1} +   \gridElem{| \grid |}}{2},  \gridElem{| \grid |} \right]$
		\Statex \Comment{Halve grid spacing}
		\State ${f} \gets \pi\unnormInd(\grid)$
		\State $F_{\textrm{new}} \gets \text{Trapezoid}(\grid, f)$
		\EndWhile \\
		\Return $\grid, {f}/F_{\textrm{new}}$ \Comment{Normalize ${f}$ to integrate to 1}
	\end{algorithmic}
\end{algorithm}
\spacingset{1.75}

Algorithm~\ref{alg:int} provides pseudocode for the adaptive integrator, as discussed in Section~\ref{subsec:int}, for numerically evaluating the \cdf{}.
We use $\grid$ to denote the grid of $\loggscale$ values, on which to evaluate the unnormalized density $\pi\unnormInd$.
Since the size of $\grid$ is changing per iteration, we denote its size by $| \grid |$ and its final element by $\gridElem{| \grid |}$.
In practice, we evaluate the log density for numerical stability and covert it to the original scale via
\[ f = \exp\Bigl[ \log \pi\unnormInd(\grid) - \max\!\left\{ \log \pi\unnormInd(\grid) \right\} \Bigr] \]
%accounting for potential changes in $\max\!\left\{ \log \pi\unnormInd(\grid) \right\} $ between iterations when evaluating relative difference between $F_{\textrm{new}}$ and  $F_{\textrm{old}}$.
In our examples, we use $\integChangeTol=0.001, \densityChangeTol=0.0001, \Delta=1$, and initialize the evaluation grid at the most recent value of $\loggscale$.

\newcommand{\designFull}{X_{\scriptscriptstyle +}}
\newcommand{\designExInt}{X}

\section{Handling estimation of intercept without shrinkage}
\label{supp:intercept}

Here we describe how to extend our spectral collapsing technique to the setting, as discussed in the last paragraph of Section~\ref{subsec:spectral_collapsing_n_bigger}, in which the intercept is given a Gaussian prior separate from the shrinkage prior on the other coefficients.
Specifically, we consider the setting in which the intercept is given a prior $\beta_0 \sim N(0, \sigma_0^2)$ and the joint prior on $\beta$ is given by
\[
\beta \mid \gscale, \lscale, \sigma_0 \sim N(0, \Sigma), \quad \Sigma = \begin{bmatrix} \sigma_0^2 & 0\\ 0 & \gscale^2 \Lscale^2 \end{bmatrix}.  
\]
To describe the necessary modifications for the fast per-$\gscale$ evaluation of $\pi\unnormInd$, we denote the full design matrix including the intercept by $\designFull$ and the matrix of the covariates excluding the intercept by $\designExInt$; i.e.\ $\designFull = [1\ \, \designExInt]$.
Correspondingly, we denote by $M_{\gscale  {\scriptscriptstyle +}}$ the matrix as defined in Equation~\eqref{eq:pn_M} and \eqref{eq:logreg_pn_M} but with $\designFull$ in place of $\designExInt$.

We present the relevant linear algebra tricks while focusing on the logistic case, in which explicit inclusion of the intercept is essential;
the linear case can be handled analogously, but it is also common to omit an intercept after centering the outcome and covariates, though such omission can technically affect inference in the presence of a prior or penalty. 
As we will see, the linear algebra manipulations make much of the computations identical to the no-intercept case.
The technique straightforwardly extends to settings in which a subset of covariates are given a separate Gaussian prior and estimated without shrinkage.

\subsection{Woodbury for $\boldsymbol{p>n}$ regime}
When treating the intercept in the same manner as all the other predictors,
%When estimating all the coefficients via a shrinkage prior with shared $\gscale$, 
the computational bottleneck in evaluating the collapsed density in Equation~\eqref{eq:logreg_cond_posterior} consisted of forming $M_\gscale = \Omega^{-1} + \gscale^2X\Lscale^2 X^\transpose$ and evaluating $|M_\gscale|$ and $z^\transpose M_\gscale^{-1} z$. 
With the separate prior on the intercept, we now must evaluate $|M_{\gscale  {\scriptscriptstyle +}}|$ and $z^\transpose M_{\gscale  {\scriptscriptstyle +}}^{-1} z$, where
\begin{align*}
M_{\gscale  {\scriptscriptstyle +}}&=\Omega^{-1} + \designFull\Sigma \designFull^\transpose\\
&=\Omega^{-1} + 
\begin{bmatrix}1 & \designExInt\end{bmatrix}  
\begin{bmatrix} \sigma_0^2 & 0\\ 0 & \gscale^2 \Lscale^2 \end{bmatrix} 
\begin{bmatrix}1^\transpose \\  \designExInt^\transpose\end{bmatrix} \\
&= \Omega^{-1} + \sigma_0^2 11^\transpose + \tau^2 \designExInt \Lscale^2 \designExInt^\transpose\\
%&= \sigma_0^2 11^\transpose + \Omega^{-1/2}(I_n + \gscale^2\Omega^{1/2} \designExInt \Lscale^2 \designExInt^\transpose\Omega^{1/2})\Omega^{-1/2}
&=  \sigma_0^2 11^\transpose + M_\gscale.
\end{align*}
The determinant  of $M_{\gscale  {\scriptscriptstyle +}}$ can be computed using the matrix determinant lemma
\begin{align*}
	\left|M_{\gscale  {\scriptscriptstyle +}} \right| = \left|\sigma_0^2 11^\transpose + M_\gscale\right| 
	&= (1+ \sigma_0^21^\transpose  M_\gscale^{-1}  1)\left| M_\gscale\right|.
\end{align*}
As for $z^\transpose M_{\gscale  {\scriptscriptstyle +}}^{-1}z $, we can apply the Woodbury formula to obtain
\begin{align*}
	z^\transpose (\sigma_0^2 11^\transpose + M_\gscale)^{-1} z 
	& = z^\transpose \{M_\gscale^{-1} - M_\gscale^{-1} 1 (\sigma_0^{-2} + 1^\transpose M_\gscale^{-1} 1)^{-1} 1^\transpose M_\gscale^{-1}\} z\\
	&= z^\transpose M_\gscale^{-1}z - \frac{(1^\transpose M_\gscale^{-1}z)^2}{\sigma_0^{-2}+1^\transpose M_\gscale^{-1} 1}.
\end{align*}
The two equations above reduce the task of evaluating $|M_{\gscale  {\scriptscriptstyle +}}|$ and $z^\transpose M_{\gscale  {\scriptscriptstyle +}}^{-1} z$ to that of evaluating $|M_\gscale|$ and bilinear forms of $M_\gscale^{-1}$, for which we have already developed fast per-$\gscale$ evaluation techniques in Equations~\eqref{eq:logreg_pn_M}, \eqref{eq:logreg_pn_det}, and \eqref{eq:logreg_pn_quad} through the spectral decomposition $\Omega^{1/2} \designExInt \Lscale^2 \designExInt^\transpose\Omega^{1/2} = VD_nV^\transpose$.

\subsection{Block operations for $\boldsymbol{n>p}$ regime}
When $n>p$, we again apply a Woodbury identity first, as in Equation~\eqref{eq:np_woodbury}, so that the required inverse is of size $p\times p$ instead of $n \times n$:
\[
		M_{\gscale  {\scriptscriptstyle +}}  ^{-1} = (\Omega^{-1} +  \designFull \Sigma \designFull^\transpose)^{-1} \\
		= \Omega - \Omega \designFull (\Sigma^{-1} +  \designFull^\transpose \Omega \designFull)^{-1} \designFull^\transpose \Omega.
\]
We can then write $\Sigma^{-1} + \designFull^\transpose \Omega \designFull$ in block form as
\begin{align*}
\Sigma^{-1} + \designFull^\transpose \Omega \designFull &=   \begin{bmatrix} \sigma_0^{-2} & 0\\ 0 & \gscale^{-2} \Lscale^{-2} \end{bmatrix} + 
\begin{bmatrix}1^\transpose \\  \designExInt^\transpose\end{bmatrix} \Omega \begin{bmatrix}1 &  \designExInt\end{bmatrix} \\
&= \begin{bmatrix}
	\sigma^{-2}_0 + 1^\transpose \Omega 1 & 1^\transpose \Omega \designExInt\\
	\designExInt^\transpose \Omega 1 & \gscale^{-2}\Lscale^{-2} + \designExInt^\transpose \Omega \designExInt 
\end{bmatrix}.
\end{align*}
By applying block matrix formulas to the above, computing the determinant and inverse of $M_{\gscale  {\scriptscriptstyle +}}$ effectively reduces to computing those of the lower right block $\gscale^{-2}\Lscale^{-2} + \designExInt^\transpose \Omega \designExInt =\gscale^{-2}\Lscale^{-1}(I_p+\gscale^{2}\Lscale \designExInt^\transpose \Omega \designExInt \Lscale)\Lscale^{-1}$, which is the same quantity we dealt with in Equation~\eqref{eq:logreg_np_woodbury}.
%In computing the determinant and inverse of $M_\tau$ via block formulas, the main expenses come from evaluating the determinant and inverse of the lower right block $\gscale^{-2}\Lscale^{-2} + \designExInt^\transpose \Omega \designExInt =\gscale^{-2}\Lscale^{-1}(I_p+\gscale^{2}\Lscale \designExInt^\transpose \Omega \designExInt \Lscale)\Lscale^{-1}$, which is the same quantity we dealt with in Equation~\eqref{eq:logreg_np_woodbury}.
We can therefore use the same spectral decomposition $\Lscale \designExInt^\transpose \Omega \designExInt \Lscale = VD_pV^\transpose$ and computations analogous to Equation~\eqref{eq:logreg_np_det} and \eqref{eq:logreg_np_quad} for fast per-$\gscale$ evaluation.

\end{document}